\documentclass{pasj00}

\begin{document}
\SetRunningHead{Y. K. Choi et al.}{Distance to VY Canis Majoris with VERA}
\Received{2008/04/21}
\Accepted{2008/08/05}

\title{Distance to VY Canis Majoris with VERA}

\author{
Yoon Kyung \textsc{Choi},\altaffilmark{1,2} \thanks{Present address is Max-Planck-Institut f${\ddot {\textrm u}}$r Radioastronomie,
Auf dem H${\ddot {\textrm u}}$gel 69, 53121 Bonn, Germany.}
Tomoya \textsc{Hirota},\altaffilmark{2,3} Mareki \textsc{Honma},\altaffilmark{2,3}
Hideyuki \textsc{Kobayashi},\altaffilmark{1,2,4,5} Takeshi \textsc{Bushimata},\altaffilmark{2,4}
Hiroshi \textsc{Imai},\altaffilmark{6} Kenzaburo \textsc{Iwadate},\altaffilmark{5} Takaaki \textsc{Jike},\altaffilmark{5}
Seiji \textsc{Kameno},\altaffilmark{6}
Osamu \textsc{Kameya},\altaffilmark{3,5} Ryuichi \textsc{Kamohara},\altaffilmark{2}
Yukitoshi \textsc{Kan-ya},\altaffilmark{7}
Noriyuki \textsc{Kawaguchi},\altaffilmark{2,3,4} Masachika \textsc{Kijima},\altaffilmark{2,3}
Mi Kyoung \textsc{Kim},\altaffilmark{1,2} Seisuke \textsc{Kuji},\altaffilmark{5}
Tomoharu \textsc{Kurayama},\altaffilmark{2} Seiji \textsc{Manabe},\altaffilmark{3,5} Kenta \textsc{Maruyama},\altaffilmark{8}
Makoto \textsc{Matsui},\altaffilmark{8}  Naoko \textsc{Matsumoto},\altaffilmark{8}
Takeshi \textsc{Miyaji},\altaffilmark{2,4}
Takumi \textsc{Nagayama},\altaffilmark{8} Akiharu \textsc{Nakagawa},\altaffilmark{6}  Kayoko \textsc{Nakamura},\altaffilmark{8}
Chung Sik \textsc{Oh},\altaffilmark{1,2}
Toshihiro \textsc{Omodaka},\altaffilmark{6}
Tomoaki \textsc{Oyama},\altaffilmark{2} Satoshi \textsc{Sakai},\altaffilmark{5} Tetsuo \textsc{Sasao},\altaffilmark{9,10}
Katsuhisa \textsc{Sato},\altaffilmark{5} Mayumi \textsc{Sato},\altaffilmark{1,2}
Katsunori M. \textsc{Shibata},\altaffilmark{2,3,4}
Yoshiaki \textsc{Tamura},\altaffilmark{3,5}  Miyuki \textsc{Tsushima},\altaffilmark{8} and Kazuyoshi \textsc{Yamashita}\altaffilmark{2,3}
        }
\altaffiltext{1}{Department of Astronomy, Graduate School of Science, The University of Tokyo, 7--3--1 Hongo, Bunkyo-ku, Tokyo 113--0033}
\altaffiltext{2}{Mizusawa VERA Observatory, National Astronomical Observatory of Japan, 2--21--1 Osawa, Mitaka, Tokyo 181--8588}
\altaffiltext{3}{Department of Astronomical Sciences, Graduate University for Advanced Studies, 2--21--1 Osawa, Mitaka, Tokyo 181--8588}
\altaffiltext{4}{Space VLBI Project, National Astronomical Observatory of Japan, 2--21--1 Osawa, Mitaka, Tokyo 181--8588}
\altaffiltext{5}{Mizusawa VERA Observatory, National Astronomical Observatory of Japan, \\ 2--12 Hoshi-ga-oka, Mizusawa-ku, Oshu-shi, Iwate 023--0861}
\altaffiltext{6}{Faculty of Science, Kagoshima University, 1--21--35 Korimoto, Kagoshima, Kagoshima 890--0065}
\altaffiltext{7}{Department of Astronomy, Yonsei University, 134 Shinchon-dong, Seodaemun-gu, Seoul 120--749, Republic of Korea}
\altaffiltext{8}{Graduate School of Science and Engineering, Kagoshima University, 1--21--35 Korimoto, Kagoshima, Kagoshima 890--0065}
\altaffiltext{9}{Department of Space Survey and Information Technology, Ajou University, Suwon 443--749, Republic of Korea}
\altaffiltext{10}{Korean VLBI Network, Korea Astronomy and Space Science Institute, P.O.Box, Yonsei University, \\ 134 Shinchon-dong, Seodaemun-gu, Seoul 120--749, Republic of Korea}

\email{ykchoi@mpifr-bonn.mpg.de}


\maketitle

\begin{abstract}
We report astrometric observations of H$_2$O masers around the red supergiant VY Canis Majoris (VY CMa) carried out with VLBI Exploration of Radio Astrometry (VERA). Based on astrometric monitoring  for 13 months, we successfully measured a trigonometric parallax of 0.88 $\pm$ 0.08 mas, corresponding to a distance of 1.14 $^{+0.11} _{-0.09}$ kpc. This is the most accurate distance to VY CMa and the first one based on an annual parallax measurement. The luminosity of VY CMa has been overestimated due to a previously accepted distance. With our result, we re-estimate the luminosity of VY CMa to be (3 $\pm$ 0.5) $\times$ 10$^5$ L$_{\odot}$ using the bolometric flux integrated over optical and IR wavelengths. This improved luminosity value makes location of VY CMa on the Hertzsprung-Russel (HR) diagram much closer to the theoretically allowable zone (i.e. the left side of the Hayashi track) than previous ones, though uncertainty in the effective temperature of the stellar surface still does not permit us to make a final conclusion.
\end{abstract}


\section{Introduction}
Massive stars play an important role in the evolution of the Universe particularly in the late stages of their evolution. Through their strong stellar winds and supernova explosions, they inject mechanical energy into the interstellar medium (ISM) \citep{bib:Abbott1982}. Also, they are principal sources of heavy elements in the ISM. In spite of their importance, our understanding of late stages of massive stars' evolution is still poor. One of the reasons for this is that massive stars are extremely rare partly because of their short lifetime. Due to their small numbers, the properties of evolved massive stars are still uncertain. For instance, as discussed in \citet{bib:Massey2003}, \citet{bib:MasseyOlsen2003} and \citet{bib:Levesque2005}, there was a discrepancy between observed and theoretically predicted locations of red supergiants, high-mass evolved stars, on the Hertzsprung-Russel (HR) diagram. Compared with stellar evolutionary models, observed red supergiants appear to be too cool and too luminous.

VY Canis Majoris (VY CMa) is one of the most well-studied red supergiants. Similar to other red supergiants, the location of VY CMa on the HR diagram is also uncertain. For instance, \citet{bib:Monnier1999}, \citet{bib:Smith2001} and \citet{bib:Humphreys2007} obtained the luminosity of (2--5) $\times$ 10$^5$ L$_{\odot}$ from the spectral energy distribution (SED) assuming the distance of 1.5 kpc \citep{bib:Lada1978} and effective temperature of 2800--3000 K based on the stellar spectral type \citep{bib:LeSidaner1996}. However, \citet{bib:Massey2006} suggested that the above parameters would make VY CMa cooler and more luminous than what current evolutionary models allow and would place it in the ``forbidden zone" of the HR diagram, which is on the right-hand side of the Hayashi track in the HR diagram (see figure \ref{fig:HRDthis}). \citet{bib:Massey2006} reexamined the effective temperature and obtained a new value of 3650 $\pm$ 25 K based on optical spectrophotometry combined with a stellar atmosphere model and suggested that the luminosity of VY CMa is only 6.0 $\times$ 10$^4$ L$_{\odot}$, which is probably a lower limit \citep{bib:Massey2008}.

\citet{bib:Levesque2005} determined the effective temperatures of 74 Galactic red supergiants based on optical spectrophotometry and stellar atmosphere models. They obtained the effective temperatures of the red supergiants with a precision of 50 K. Their new effective temperatures are warmer than those in the literature. 
These new effective temperature values seem to be consistent with the theoretical stellar evolutionary tracks. However, the luminosities of the red supergiants still had large uncertainty due to possible errors in estimated distances. Since most of red supergiants are very far, it has been difficult to apply the most reliable trigonometric parallax method to their distance measurements.

In the case of VY CMa, the currently accepted distance of 1.5 kpc \citep{bib:Lada1978} is obtained by assuming that VY CMa is a member of the NGC 2362 star cluster and that the distance of NGC 2362 is same as that of VY CMa. The distance to the NGC 2362 was determined from the color-magnitude diagram \citep{bib:Johnson1961} with an accuracy of no better than 30 \%. Since the luminosity depends on the square of the distance, the accuracy of luminosity estimation is worse than 50 \%. The proper motions measured with H$_2$O masers in previous studies (\cite{bib:Richards1998}; \cite{bib:Marvel1998}) have not been contradictory to the above distance value (1.5 kpc). When the H$_2$O maser region around a star is modeled with a symmetrically expanding spherical shell, the distance to the star from the Sun can be inferred by assuming that the observed mean proper motion multiplied by the distance should be equal to the observed mean 
radial velocity of the H$_2$O masers. The proper motion velocities in \citet{bib:Richards1998} are well consistent with the H$_2$O maser spectral line velocities at the distance of 1.5 kpc. Also, \citet{bib:Marvel1998} derived the distance of VY CMa to be 1.4 $\pm$ 0.2 kpc. However, such a ``statistical parallax'' method depends on kinematical models fitted to the observed relative proper motions. In fact, if we add rotation or anisotropic expansion to the simple spherical expansion model, the inferred distance value could vary beyond quoted error ranges. An accurate distance determination without an assumption is crucial for obtaining the right location of VY CMa on the HR diagram and for determining fundamental parameters of the star.

Thanks to the recent progress in the VLBI technique, the distance measurements with trigonometric parallaxes have become possible even beyond 5 kpc (e.g., \cite{bib:Honma2007}). Since VY CMa has strong maser emission in its circumstellar envelope, we have conducted an astrometric observations of H$_2$O masers around VY CMa to measure an accurate parallax with VLBI Exploration of Radio Astrometry (VERA) and here we present the results.

\section{Observations and Data Reduction}

We have observed H$_2$O masers (H$_2$O 6$_{16}$--5$_{23}$ transition at the rest frequency of 22.235080 GHz) in the red supergiant VY CMa with VERA at 10 epochs over 13 months. The epochs are April 24, May 24, September 2, October 30, November 27 in 2006, January 10, February 14, March 26, April 21 and May 27 in 2007 (day of year 114, 144, 245, 303, 331 in 2006, 010, 045, 085, 111 and 147 in 2007, respectively). In each epoch, VY CMa and a position reference source J0725--2640 [$\alpha$(J2000) = 07h25m24.413135s, $\delta$(J2000) = --26d40'32.67907" in VCS 5 catalog, \citep{bib:Kovalev2007}] were observed simultaneously in dual-beam mode for about 7 hours. The separation angle between VY CMa and J0725--2640 is 1.059 degrees. The instrumental phase difference between the two beams was measured at each station during the observations based on the correlations of artificial noise sources (\cite{bib:Kawaguchi2000}; \cite{bib:Honma2008b}). A bright continuum source (DA193 at 1st--3rd epochs and J0530+1330 at other epochs) was observed every 80 minutes for bandpass and delay calibration at each beam.

Left-handed circularly polarized signals were sampled with 2-bit quantization and filtered with the VERA digital filter unit \citep{bib;Iguchi2005}. The data were recorded onto magnetic tapes at a rate of 1024 Mbps, providing a total bandwidth of 256 MHz which consists of 16 $\times$ 16 MHz IF channels. One IF channel was assigned to the H$_2$O masers in VY CMa and the other 15 IF channels were assigned to J0725--2640,  respectively. The correlation processing was carried out on the Mitaka FX correlator \citep{bib:Chikada1991}. The spectral resolution for H$_2$O maser lines is 15.625 kHz, corresponding to the velocity resolution of 0.21 km s$^{-1}$.

All data reduction was conducted using the NRAO Astronomical Image Processing System (AIPS) package. The amplitude and the bandpass calibration for target source (VY CMa) and reference source (J0725--2640) were performed independently. The amplitude calibration of each antenna was performed using system temperatures measured during observation. The bandpass calibration was applied using a bright continuum source (DA193 at 1st--3rd epochs and J0530+1330 at other epochs). Doppler corrections were carried out to obtain the radial motions of the H$_2$O masers relative to the Local Standard of Rest (LSR).

Then, we calibrated the clock parameters using the residual delay of a bright continuum calibrator, which is also used for the bandpass calibration. The fringe fitting was made on the position reference source J0725--2640 with integration time of 2 minutes and time interval of 12 seconds to obtain residual delays, rates, and phases. These phase solutions were applied to the target source VY CMa and we also applied the dual-beam phase calibration \citep{bib:Honma2008b} to correct the instrumental delay difference between the two beams. We calibrated the tropospheric zenith delay offset by using the GPS measurements of the tropospheric zenith delay made at each station \citep{bib:Honma2008a}. After these calibrations, synthesized clean images were obtained. A typical synthesized beam size (FWHM) was 2 mas $\times$ 1 mas with a position angle of --23$^{\circ}$. We measured positions of the H$_2$O maser features relative to the extragalactic source J0725--2640 with two-dimensional Gaussian fitting.

\section{Results}
\begin{figure}
  \begin{center}
    \FigureFile(75mm,75mm){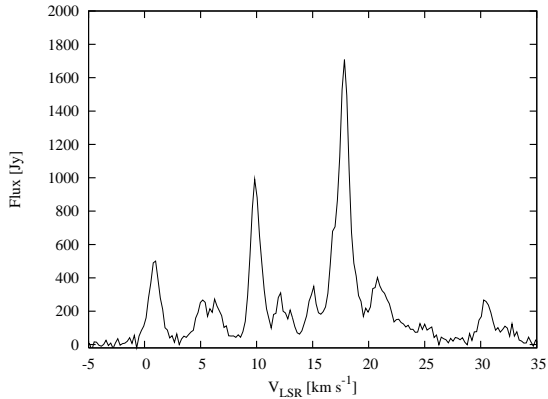}
  \end{center}
  \caption{The total power spectrum of the H$_2$O masers in VY CMa obtained at Iriki station on April 24, 2006.}\label{fig:Kspectrum}
\end{figure}
The autocorrelation spectrum of the H$_2$O masers around VY CMa is shown in figure \ref{fig:Kspectrum}. The spectrum shows rich maser emission over LSR velocities ranging from --5 to 35 km s$^{-1}$. Though there are variations in flux density, overall structures of maser spectrum are common to all the epochs, indicating that maser spots are surviving over the observing period of 13 months.

\begin{figure*}
  \begin{center}
    \FigureFile(100mm,100mm){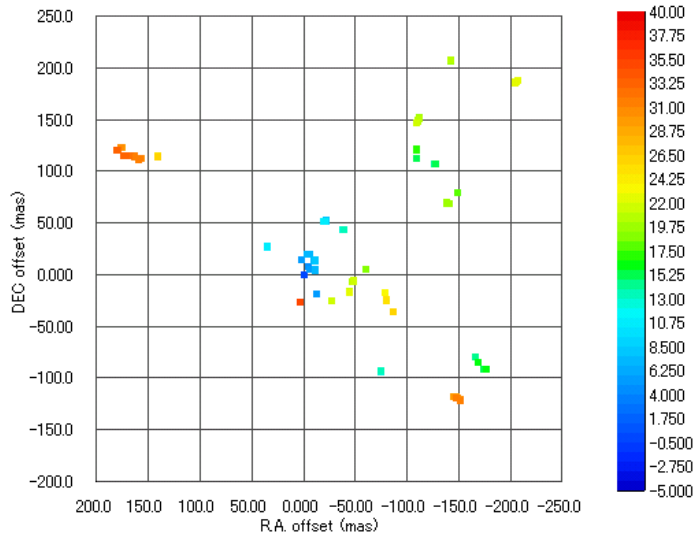}
  \end{center}
  \caption{The distribution of the H$_2$O masers in VY CMa obtained on April 24, 2006. Color, as shown on the scale on the right-hand side, represents the LSR velocities of the maser features.}\label{fig:Kmap}
\end{figure*}
To reveal the distribution of the H$_2$O masers, we mapped the H$_2$O maser features in VY CMa at the first epoch (April 24, 2006). We detected 55 maser features that have signal-to-noise ratio larger than 7 in two adjacent channels. The distribution of the H$_2$O maser features is shown in figure \ref{fig:Kmap}. The most red-shifted component and the most blue-shifted component lie near the apparent center of the distribution and the moderately red-shifted components on a NE-SW line of 150 mas. This distribution well agrees with the previous observational results by \citet{bib:MarvelPhD}.

Among the H$_2$O maser features in VY CMa, one maser feature at a LSR velocity of about 0.55 km s$^{-1}$ is analyzed to detect an annual parallax. This feature, the most blue-shifted discrete component, is stably detected at all epochs. It is the third brightest maser component in the total-power spectrum, but in fact it is the second brightest component in the map. We note that while the brightest channel has multiple components, there is only one component in the channel at the LSR velocity of about 0.55 km s$^{-1}$. Because of the simple structure as well as its strength, this velocity channel is most suitable for astrometric measurements. Analyses for the other H$_2$O maser features will be reported in a forthcoming paper.

\begin{figure*}
  \begin{center}
    \begin{tabular}{cc}
      \FigureFile(85mm,85mm){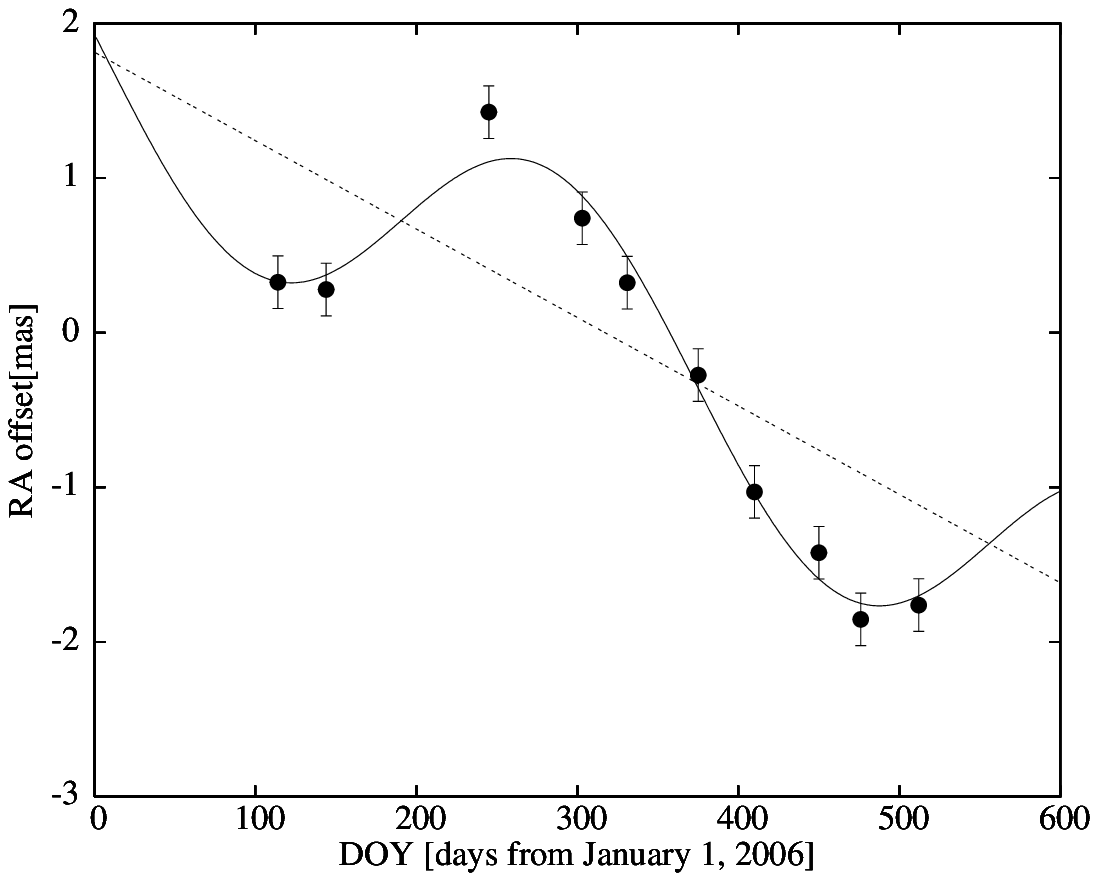} &
      \FigureFile(85mm,85mm){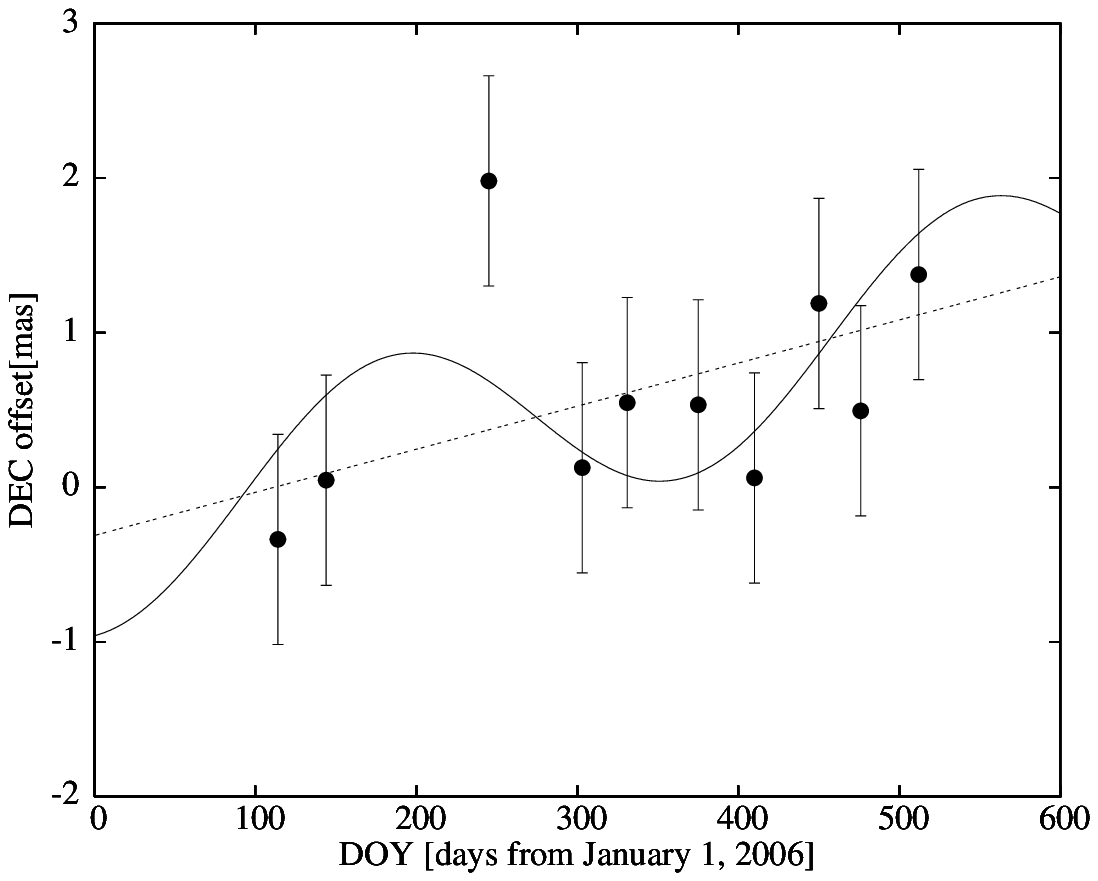} \\
    \end{tabular}
  \caption{Results of the measured positions of the H$_2$O maser spot at the LSR velocity of 0.55 km s$^{-1}$ in VY CMa using J0725--2640 as a position reference source. The position offsets are with respect to $\alpha$(J2000.0) = ${\rm 07^h 22^m 58^s.32906}$, $\delta$(J2000.0) = --25$^{\circ}$46'03".1410. \textit{The left panel} shows the movements of the maser spot in right ascension as a function of time (day of year). \textit{The right panel} is the same as \textit{the left panel} in declination. Solid lines represent the best fit model with an annual parallax and a linear proper motion for the maser spot. Dotted lines represent the linear proper motion (--2.09 $\pm$ 0.16 mas yr$^{-1}$ in right ascension and 1.02 $\pm$ 0.61 mas yr$^{-1}$ in declination) and points represent observed positions of maser spot with error bars indicating the positional uncertainties in systematic errors (0.17 mas in right ascension and 0.68 mas in declination).}\label{fig:para}	
  \end{center}
\end{figure*}

Figure \ref{fig:para} shows the position measurements of the 0.55 km s$^{-1}$ H$_2$O maser component for 13 months. The position offsets are with respect to $\alpha$(J2000.0) = ${\rm 07^h 22^m 58^s.32906}$, $\delta$(J2000.0) = --25$^{\circ}$46'03".1410. Assuming that the movements of maser features are composed of a linear motion and the annual parallax, we obtained a proper motion and an annual parallax by the least-square analysis. The declination data have too large errors to be suitable for an annual parallax and a proper motion measurement. Therefore, we obtained the parallax of VY CMa to be 0.88 $\pm$ 0.08 mas, corresponding to a distance of 1.14 $^{+0.11} _{-0.09}$ kpc using only the data in right ascension. This is the first distance measurement of VY CMa based on an annual parallax measurement with the highest precision. We estimated the positional uncertainties from the least-square analysis, and the values of the errors are 0.17 mas in right ascension and 0.68 mas in declination making reduced $\chi^{2}$ to be 1.

We also determined the absolute proper motion in right ascension and declination. The absolute proper motion is --2.09 $\pm$ 0.16 mas yr$^{-1}$ in right ascension and 1.02 $\pm$ 0.61 mas yr$^{-1}$ in declination. Compared with the proper motion of VY CMa obtained with Hipparcos, 9.84 $\pm$ 3.26 mas yr$^{-1}$ in right ascension and 0.75 $\pm$ 1.47 mas yr$^{-1}$ in declination \citep{bib:Perryman1997}, there is a discrepancy in the proper motion in right ascension. Nearly 12 mas yr$^{-1}$ difference (more than 65 km s$^{-1}$ at 1.14 kpc) seems too large even when we take into account possible relative motion of the 0.55 km s$^{-1}$ maser feature with respect to the star itself. The difference may be due to the fact that circumstellar envelope of VY CMa has complex small-scale structure at optical wavelengths. This might seriously affect proper motion measurements with Hipparcos.

The quantitative estimation of the individual error sources in the VLBI astrometry is difficult as previously mentioned in \citet{bib:Hachisuka}, \citet{bib:Honma2007}, and \citet{bib:Hirota2007}. Therefore, we estimated errors from the standard deviations of the least-square analysis to be 0.17 mas in right ascension and 0.68 mas in declination. When we consider that the statistical errors of the position, 0.02--0.09 mas, are estimated from residuals of the Gaussian fitting, the large values of the standard deviations suggest that some systematic errors affect the result of astrometry. In the following, we discuss causes of these astrometric errors, although we cannot separate each factor of error.

First, we consider the errors originating from the reference source. The positional errors of the reference source J0725--2640 affect those of the target source. The position of J0725--2640 was determined with an accuracy of 0.34 mas in right ascension  and 0.94 mas in declination, respectively \citep{bib:Kovalev2007}. Because these offsets are constant at all epochs, it dose not affect the parallax measurements. Also, when the reference source is not a point source, the positional errors of the target source could occur due to the structure and its variation of the reference source. However, since the reference source for our measurements is point-like and shows no structural variations between epochs, this is not likely to be the main source of the positional errors.

Secondly, we consider the baseline errors originated from the positional errors of each VLBI station. The positions of VERA stations are determined with an accuracy of 3 mm by the geodetic observations at 2 and 8 GHz every 2 weeks. The positional errors derived from the baseline errors are 11 $\mu$as at a baseline of 1000 km with the baseline error of 3 mm. This error is much smaller than our astrometric errors.

Thirdly, a variation of the structure of the maser feature could be one of the error sources. \citet{bib:Hirota2007}, \citet{bib:Imai2007}, and \citet{bib:Hirota2008} proposed the maser structure effect as main sources in their trigonometric parallax measurements for nearby star forming regions Orion KL, IRAS 16293--2422 in $\rho$ Oph East, and SVS 13 in NGC 1333, correspondingly. However, the effect does not seem dominating in our case because, (1) the 0.55 km s$^{-1}$ maser feature showed a stable structure in the closure phase, spectrum and map at all epochs of our observation, (2) this effect is inversely proportional to the distance of the target source and hence should be more than twice less significant for VY CMa as those in the above cases for a given size of the structure variation, (3) it is difficult to explain the large difference between astrometric errors in right ascension and declination by this effect.

Finally, we have to consider the errors by the zenith delay residual due to tropospheric water vapor. These errors originate from the difference of path length through the atmosphere between the target and the reference sources, and generally larger in declination than in right ascension. According to the result of the simulation in \citet{bib:Honma2008a}, the positional error by the tropospheric delay is 678 $\mu$as in declination when the atmospheric zenith residual is 3 cm, the declination is --30 degrees, separation angle (SA) is 1 degree, and the position angle (PA) is 0 degrees. This is well consistent with our measurements. Therefore, the atmospheric zenith delay residual is likely to be the major source of the astrometric errors.

\section{The Location on the HR Diagram}
We successfully detected a trigonometric parallax of 0.88 $\pm$ 0.08 mas, corresponding to a distance of 1.14 $^{+0.11} _{-0.09}$ kpc to VY CMa. Compared with the previously accepted distance 1.5 kpc \citep{bib:Lada1978}, the distance to VY CMa became 76 \%. Since the luminosity depends on the square of the distance, the luminosity should become 58 \% of previous estimates. Hence, here we re-estimate the luminosity of VY CMa with the most accurate distance.
The luminosity can be estimated as follows:
\begin{equation}
L = 4 \pi  d^2  F_{bol},
\end{equation}
where $L$ is luminosity, $d$ is distance and $F_{bol}$ is the bolometric flux. To obtain F$_{bol}$, we used the SED of VY CMa. The data are based on HST optical images and near-IR ground based images in \citet{bib:Smith2001} and IRAS fluxes from 25 to 100 $\mu$m. The F$_{bol}$ is obtained by integrating the observed fluxes. The estimated luminosity of VY CMa with our distance is (3 $\pm$ 0.5) $\times$ 10$^5$ L$_{\odot}$.

We re-estimate the luminosities of VY CMa in the previous studies with our distance. \citet{bib:LeSidaner1996} obtained a luminosity of VY CMa to be 9 $\times$ 10$^5$ L$_{\odot}$ from the SED at a kinematic distance of 2.1 kpc. With the distance of 1.14 kpc, the luminosity of \citet{bib:LeSidaner1996} became 2.6 $\times$ 10$^5$ L$_{\odot}$. \citet{bib:Smith2001} also estimated a luminosity of VY CMa from the SED and their luminosity is 5 $\times$ 10$^5$ L$_{\odot}$ at a distance of 1.5 kpc, which is is revised to be 3.0 $\times$ 10$^5$ L$_{\odot}$ using our distance. These luminosities are well consistent with each other. \citet{bib:Massey2006} estimated an effective temperature of VY CMa to be 3650 K, which fitted the MARCS atmosphere model to the observed spectrophotometric data. They estimated an absolute V magnitude $M_V$ using the observed V magnitude, currently known distance of 1.5 kpc and a visible extinction $A_V$. Since the MARCS model presents bolometric corrections as a function of effective temperature, they obtained the luminosity of VY CMa to be 6.0 $\times$ 10$^4$ L$_{\odot}$, much lower than our value derived above. When they adopt our distance, their luminosity would be increased. The accurate distance is essential to estimate the accurate luminosity.  

To place VY CMa on the HR diagram, we adopted an effective temperature from the literatures. As we already mentioned, \citet{bib:Massey2006} estimated an effective temperature of VY CMa to be 3650 K based on the observed spectrophotometric data. With the previously accepted spectral type of M4-M5, \citet{bib:LeSidaner1996} obtained an effective temperature of 2800 K and \citet{bib:Smith2001} also adopted an effective temperature of 3000 K. Since the effective temperature does not depend on the distance, our measurements cannot judge which effective temperature is correct. 

\begin{figure*}
  \begin{center}
    \FigureFile(100mm,100mm){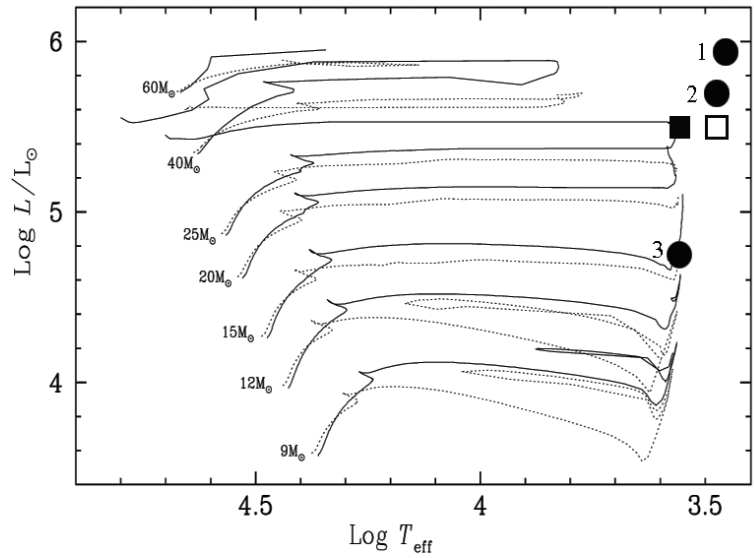}
  \end{center}
  \caption{The various locations of VY CMa on the HR diagram. The filled and the open squares represent our results. The luminosity of our result is calculated using the distance based on the trigonometric parallax measurements and the bolometric flux from the SED. The filled square adopted the effective temperature of 3650 K \citep{bib:Massey2006} and the open square adopted 3000 K\citep{bib:Smith2001}. The circles ``1,''  ``2''  and ``3'' represent the results of \citet{bib:LeSidaner1996}, \citet{bib:Smith2001} and \citet{bib:Massey2006}, respectively. The evolutionary tracks are from \citet{bib:Meynet2003}.} \label{fig:HRDthis}
\end{figure*}

Although we cannot estimate the effective temperature of VY CMa with our measurements, when we adopt the effective temperature of 3650 K from the MARCS atmosphere model \citep{bib:Massey2006}, the location of VY CMa on the HR diagram is determined as the filled square in figure \ref{fig:HRDthis}. For comparison, we also show VY CMa's locations on the HR diagram obtained in the previous studies as filled circles. Our results suggest that the location of VY CMa on the HR diagram is now consistent with the evolutionary track of an evolved star with an initial mass of 25 M$_{\odot}$. Also, re-scaled luminosity values of \citet{bib:LeSidaner1996} and \citet{bib:Smith2001} imply much closer locations to the 25 M$_{\odot}$ track than their original ones which were deeply inside the ``forbidden zone''. On the other hand, the lower limit of luminosity of 6.0 $\times$ 10$^4$ L$_{\odot}$ in \citet{bib:Massey2006} would be consistent with 15 M$_{\odot}$ initial mass of VY CMa. According to \citet{bib:Hirschi2004}, there is an order of magnitude difference in lifetimes between 15 M$_{\odot}$ and 25 M$_{\odot}$ in the initial mass. The improvement in mass and age values should affect statistical studies on evolution of massive stars such as the initial mass function \citep{bib:Salpeter1955}.  

On the other hand, there is an argument against the effective temperature in \citet{bib:Massey2006}. \citet{bib:Humphreys2006} suggested that the spectrum in \citet{bib:Massey2006} are more like their M4-type reference spectrum than M2-type reference spectrum. \citet{bib:Humphreys2006} also pointed out that the modeling applicable for ``standard" stars without mass-loss (MARCS) is simply not valid for an object like VY CMa. When we adopt the previously accepted spectral type, instead of M2.5I \citep{bib:Massey2006}, the effective temperature is 3000 K \citep{bib:Smith2001}. The location of VY CMa on the HR diagram is shown as the open square in figure \ref{fig:HRDthis}. In this case, the position on the HR diagram is still not consistent with the theoretical evolutionary track. Table \ref{tab:first} summarized our adopted parameters for VY CMa.

\begin{table*}
  \caption{Adopted parameters for VY CMa}\label{tab:first}
  \begin{center}
    \begin{tabular}{lll}
      \hline
       Parameter & Value &  Note \\ \hline
      Parallax & 0.88 $\pm$ 0.08 mas \\
      Distance & 1.14 $^{+0.11} _{-0.09}$ kpc \\
      Luminosity & (3 $\pm$ 0.5) $\times$ 10$^5$ L$_{\odot}$  \\
      Mass & 25 M$_{\odot}$ & \citet{bib:Meynet2003} \\
      Temperature  & 3650 $\pm$ 25 K & \citet{bib:Massey2006} \\
                         & 3000 K &\citet{bib:Smith2001} \\
      \hline
    \end{tabular}
  \end{center}
\end{table*}

The accurate distance measurements of red supergiants, which provide true luminosities, will greatly contribute to the understanding of massive star evolution, though there is still uncertainty in the effective temperature on the stellar surface.  

\section{Conclusion}

We have observed the H$_2$O masers around the red supergiant VY CMa with VERA during 10 epochs spread over 13 months. Simultaneous observations for both H$_2$O masers around VY CMa and the position reference source J0725--2640 were carried out. We measured a trigonometric parallax of 0.88 $\pm$ 0.08 mas, corresponding to a distance of 1.14 $^{+0.11} _{-0.09}$ kpc from the Sun. It is the first result that the distance of VY CMa is determined with an annual parallax measurement. There had been overestimation of the luminosities in previous studies due to the previously accepted distance. Using the most accurate distance based on the trigonometric parallax measurements and the bolometric flux from the observed SED, we estimated the luminosity of VY CMa to be (3 $\pm$ 0.5) $\times$ 10$^5$ L$_{\odot}$. The accurate distance measurements provided the improved luminosity. The location of VY CMa on the HR diagram became much close to the theoretically allowable region, though there is still uncertainty in the effective temperature.  
 
~\\
We are grateful to the referee, Dr. Philip Massey, for helpful comments on the manuscript. We would like to thank Prof. Dr. Karl M. Menten for his invaluable comments and for his help improving the manuscript. The authors also would like to thank all the supporting staffs at Mizusawa VERA observatory for their assistance in observations.


\end{document}